\documentclass[conference]{IEEEtran}
\IEEEoverridecommandlockouts
\usepackage{cite}
\usepackage{amsmath,amssymb,amsfonts}
\usepackage{algorithmic}
\usepackage{graphicx}
\usepackage{textcomp}
\usepackage{enumitem}
\usepackage{subcaption}
\usepackage{comment}
\usepackage{listings}
\usepackage{color}
\usepackage{xcolor}
\usepackage{hyperref}

\definecolor{mygreen}{rgb}{0,0.6,0}
\definecolor{mygray}{rgb}{0.5,0.5,0.5}
\definecolor{mymauve}{rgb}{0.58,0,0.82}

\definecolor{backcolour}{rgb}{0.95,0.95,0.92}

\lstdefinestyle{promptstyle}{
    backgroundcolor=\color{backcolour},
}

\usepackage{array}
\newcolumntype{C}[1]{>{\centering\arraybackslash}p{#1}}

\def\BibTeX{{\rm B\kern-.05em{\sc i\kern-.025em b}\kern-.08em
    T\kern-.1667em\lower.7ex\hbox{E}\kern-.125emX}}
\begin{document}
\title{Customizing a Large Language Model for VHDL Design of High-Performance Microprocessors}

\author{
\IEEEauthorblockN{
Nicolas Dupuis\textsuperscript{\textsection}\IEEEauthorrefmark{1},
Ravi Nair\IEEEauthorrefmark{2}\IEEEauthorrefmark{1},
Shyam Ramji\textsuperscript{\textsection}\IEEEauthorrefmark{1},
Sean McClintock\IEEEauthorrefmark{2},
Nishant Chauhan\IEEEauthorrefmark{2},\\
Priyanka Nagpal\IEEEauthorrefmark{2},
Bart Blaner\IEEEauthorrefmark{2},
Ken Valk\IEEEauthorrefmark{2},
Leon Stok\IEEEauthorrefmark{2},
and Ruchir Puri\textsuperscript{\textsection}}
\IEEEauthorblockA{\textsuperscript{\textsection}IBM Research, Yorktown Heights, NY, USA}%
\IEEEauthorblockA{\IEEEauthorrefmark{2}IBM Infrastructure}%
}

\maketitle
\begingroup\renewcommand\thefootnote{\IEEEauthorrefmark{1}}
\footnotetext{Equal contribution}
\endgroup

\begin{abstract}
The use of Large Language Models (LLMs) in hardware design has taken off in recent years, principally through its incorporation in tools that increase chip designer productivity. There has been considerable discussion about the use of LLMs in RTL specifications of chip designs, for which the two most popular languages are Verilog and VHDL. LLMs and their use in Verilog design has received significant attention due to the higher popularity of the language, but little attention so far has been given to VHDL despite its continued popularity in the industry. There has also been little discussion about the unique needs of organizations that engage in high-performance processor design, and techniques to deploy AI solutions in these settings. In this paper, we describe our journey in developing a Large Language Model (LLM) specifically for the purpose of explaining VHDL code, a task that has particular importance in an organization with decades of experience and assets in high-performance processor design. We show how we developed test sets specific to our needs and used them for evaluating models as we performed extended pretraining (EPT) of a base LLM. Expert evaluation of the code explanations produced by the EPT model increased to 69\% compared to a base model rating of 43\%. We further show how we developed an LLM-as-a-judge to gauge models similar to expert evaluators. This led us to deriving and evaluating a host of new models, including an instruction-tuned version of the EPT model with an expected expert evaluator rating of 71\%. Our experiments also indicate that with the potential use of newer base models, this rating can be pushed to 85\% and beyond. We conclude with a discussion on further improving the quality of hardware design LLMs using exciting new developments in the Generative AI world.
\end{abstract}


\newcommand{\listingsttfamily}{\fontfamily{IBMPlexMono-TLF}\tiny}

\lstdefinestyle{prettycode}{
  basicstyle=\listingsttfamily,
  backgroundcolor=\color{backcolour},
  aboveskip={0.9\baselineskip},               
  keepspaces=true,
}
\lstset{style=prettycode}

\section{Introduction}
The complexity of modern chip design has grown considerably, driven by the increasing demand for high-performance and AI-optimized chips. As semiconductor technology advances, the design and verification of digital systems require significant engineering effort and specialized domain expertise. This challenge is exacerbated by the growing complexity of microprocessor architectures, stringent performance and power requirements, and rapid product development cycles. Generative AI and Large Language Models (LLMs) have emerged as promising technologies to supplement Electronic Design Automation (EDA) workflows by augmenting engineers' capabilities, automating repetitive tasks, and accelerating design iteration cycles \cite{ml-eda-survey}\cite{ml-cad-survey}. Researchers have demonstrated the potential of LLMs in generating scripts for EDA tools\cite{chipnemo}, in conversational LLMs for design assistance\cite{chipgpt}\cite{chipchat}\cite{chateda}, in Physical Design \cite{openroad-assistant} and High-level Synthesis\cite{c2hlsc}\cite{hls-program-repair} tasks, and in addressing hardware security\cite{llm-hw-assertions}. LLMs have shown promise also in streamlining the design process by generating Register Transfer Level (RTL) descriptions from natural language specifications ~\cite{chipnemo}\cite{chipgpt}\cite{autochip}\cite{rtllm}.

While much research has focused on LLM-aided Verilog generation\cite{rtlmcts}\cite{verilog-code-benchmark}\cite{verilog-gen}, there have been hardly any efforts on VHDL, an RTL language widely used in microprocessor designs, and in the aerospace and automotive industries. Industrial organizations designing complex microprocessors often rely on VHDL for its strong type checking, deterministic concurrency modeling, and structured syntax. However, there exists a significant gap in VHDL expertise and skills, posing challenges to productivity and design correctness. Consequently, LLM-assisted RTL design tools targeting VHDL can play a crucial role, not only in improving product quality by mitigating design risks, but also by bridging the skills gap by improving onboarding efficiency. Supplementing HDL code generation\cite{verilog-eval}, LLMs can also serve as intelligent assistants capable of explaining complex RTL constructs to aid designers, especially those new to the process, in both designing and debugging. 

The potential of LLMs in VHDL-based designs was shown in\cite{vhdl-codesc}. Practical use of LLMs in production-level performance-sensitive design settings requires further customization of LLMs with domain- and organization-specific data.

Designing an LLM assistant for the development of real-world, high-performance microprocessor  presents several challenges. Firstly, handling confidential design data is a critical concern, as proprietary microprocessor designs and specifications must be protected against unauthorized access and leakage. Ensuring data privacy while leveraging LLMs requires careful consideration of access controls, data anonymization techniques, and secure model training paradigms. Secondly, harvesting and structuring design knowledge accumulated over generations of microprocessor development is nontrivial, as specifications, documentation, and design insights are often scattered across unstructured datasets. Effective data filtering and preprocessing techniques are necessary to distill meaningful training data for LLM fine-tuning. Furthermore, training strategies and evaluation methodologies must account for the limitations of domain-specific expertise, as subject matter experts (SMEs) are limited in number and cannot be depended upon to manually label or verify large datasets. 

In this paper, we delve into our journey of customizing an LLM tailored specifically for VHDL code and design explanation – a task of utmost importance for organizations with decades of experience and a rich portfolio of high-performance processor designs. Onboarding new designers in such an environment is difficult and calls for a tool which provides accurate and contextually relevant explanations of complex hardware descriptions. Building a customized LLM for the tool not only serves as a design and educational aid but also facilitates the capture and preservation of the extensive design knowledge and expertise within the organization. This would be an invaluable resource for future design engineers. 

Toward that goal, starting with a base LLM, we show how we improved the performance significantly using extended pretraining (EPT) techniques with domain-specific unlabeled design data. We saw an improvement in the performance of our LLM, from a base model score of 43\% to a EPT model score of 69\%, and further to 71\% through model merging\cite{ties-merging}\cite{chip-align}, and instruction tuning (IT)\cite{instruct-tune}\cite{instruct-tune-eval}. We explored the concept of LLM-as-a-Judge to evaluate models aligned with SME ratings to overcome the aforementioned limitation in the number of experts to support the rapid cycles of model training and instruction-tuning. Our exploration of the LLM space, sped up with an LLM-as-a-Judge, eventually pointed us to models that push up the score to high levels (higher than 85\%), marking a significant step forward in the use of LLMs within an industrial design flow. These encouraging results will motivate further work in deploying LLM-aided assistants to improve productivity and design reliability in confidential, high-performance design environments.

This paper is organized as follows. Section \ref{Methods} discusses the various considerations in the design of our assistant and our approach to tackle the challenges encountered. This includes an overview of the secure infrastructure that we used, details about our test set generation, our extended pretraining process, our evaluation process, and our efforts on automating explanation evaluation. Section \ref{Results} presents the results of our extended pretraining and instruction tuning work, an evaluation of our LLM-as-a-judge, and some experimental projection for the future. Section \ref{Discussion} closes the paper with a discussion of our efforts and the impact of our work.

\section{Methods and materials}\label{Methods}

\subsection{Infrastructure}
Compared to general AI projects, there are several special considerations needed for an industrial project as described here. Security is the chief among them. Semiconductor chips, especially processors, are valuable intellectual property and hence need to be designed in highly secure environments. Moreover, the tools that need to be developed in support of these designs must also be tested, and often developed, in such secure environments. Unlike traditional programs, AI models cannot be developed outside a secure environment and simply ported into a secure environment. Rather, the effectiveness of the AI model requires training it on data that is proprietary and hence its development must be done in a secure environment. 

\begin{figure}[hbtp]
    \centering
    \includegraphics[width=1\linewidth]{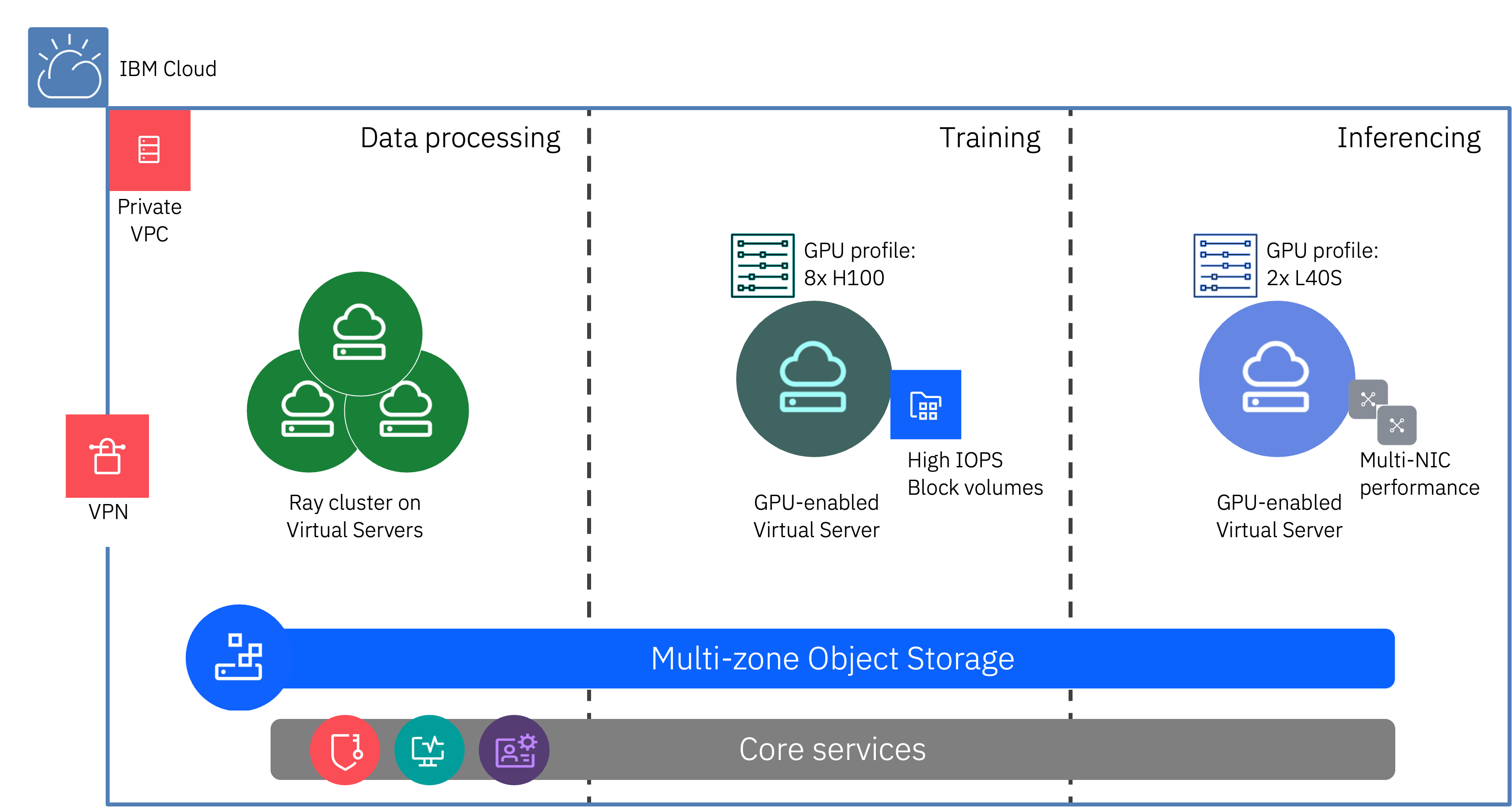}
    \caption{Overview of Project Infrastructure}
    \label{infra}
\end{figure}

Fig. \ref{infra} shows an overview of the infrastructure deployed for this project. A secure VPC was created in IBM Cloud and multiple clusters were provisioned within this VPC. Training data was transferred securely to a multi-region object storage bucket with the preprocessing cluster, the training cluster, and the inferencing cluster communicating through data shared in this bucket. The bucket itself was protected using key protection.
\subsection{Document Acquisition and Preprocessing}\label{doc_acquisition}
To help with the identification of the types of material to be fed into the extended pretraining (EPT) process we engaged subject matter experts (SMEs). The following sources were identified:
\begin{itemize}
\item GitHub projects
\item Wiki projects
\item Box folders related to the project
\item Material from educational courses and tutorials offered within the organization 
\item Manuals, processor architecture documents, related product offering documents, and papers authored by members of the organization 
\end{itemize}
All the material collected were from internal sources within the organization, 
most of which are not available publicly and hence not used for training publicly available LLM base models. The collected material was classified into the following buckets for preprocessing:
\begin{itemize}
\item Code in various languages
\item Markdown files, principally from GitHub
\item Text files, word processing documents (e.g. Word), and presentation files (e.g. PowerPoint), mainly from project repositories, such as Box
\item HTML files from the Web
\item PDF files, particularly from papers, public documents, manuals, and books
\end{itemize}

We used an open-source data processing pipeline called 
{Data prep Kit}\cite{data-prep-kit} to process most of our collected data. 

The first step in the pipeline was \textit{filtering}. In this step, we eliminated documents not pertaining to the type of knowledge we wanted the model to learn. Examples of these were 
\begin{enumerate}[label=(\alph*)]
\item code in languages that were not of direct interest, e.g. languages used for describing physical layout of circuits,
\item HTML files that only contained Web links, and 
\item slide presentations dealing with project logistics (e.g. schedule, status) rather than technical content.
\end{enumerate}

Next came the step of \textit{deduplication}. We did a pairwise comparison of contents to eliminate files that were obviously identical to others left in the set. This was followed by a scan of document names to identify same content but in different formats, for example PDF files that were direct conversions from source documents. In a future iteration, we will look for further pruning of information by using fuzzy deduplication techniques.

This was followed by a \textit{conversion} step that converted all non-text and non-code files to a markdown format. Most of the document types had converters in the kit that supported markdown as the target format. The exception was in the cases of Word and PowerPoint files where the original documents had to be converted to HTML files before getting converted to markdown.

The final step was the \textit{tokenization} of the markdown files along with the unconverted code and text files. Tokenization was done with the Starcoder tokenizer\cite{starcoder}.

\subsection{Test Generation}
Test sets form an important part of any AI project. While the distribution and quality of the test set usually get a lot of attention, the logistics of evaluating the results of the model using the test set must also be kept in mind while creating the test set. We decided to have two categories of test sets, one targeting Code Explanation, the skill that we wanted the model to acquire, and another to evaluate the knowledge acquired by the model to serve as the foundation for other skills to be added later. For the Code Explanation skill, we used subject matter experts to choose or construct a piece of code written in VHDL, or its internal proprietary variants, 
and provide reference explanations for the code. An example of such a code-explanation pair is shown in Fig. \ref{CEQuestion}. 

\begin{figure}[hbtp]
\begin{lstlisting}[language=Python]
question = '''
Please provide an expert explanation for the VHDL code snippet provided below. Your explanation should be less than 150 words.
<vhdl>
function pcoc_byte_reord_b2l (
                         data : in std_ulogic_vector
                        ) return std_ulogic_vector is

constant width      : positive := data'length;
constant num_bytes  : positive := width/8;
variable result     : std_ulogic_vector(0 to width-1);

begin
   for i in 0 to num_bytes-1 loop
      result((num_bytes-1-i)*8 downto ((num_bytes-i)*8)-1) := data(i*8 to ((i+1)*8)-1);
   end loop;

   return result;
end pcoc_byte_reord_b2l;
</vhdl>
'''
reference_explanation = '''
Reorder bytes from big endian to little endian.
'''
\end{lstlisting}
\caption{Sample Code Explanation question}\label{CEQuestion}
\end{figure}

For knowledge acquisition we had the experts construct multiple-choice questions. The instructions in the prompt template required the LLM to generate only a letter answer corresponding to the correct choice. This minimized the post-processing needed for the output from the LLM and allowed automation of the test-set evaluation. A sample Multiple-Choice question is shown in Fig. \ref{MCQuestion}. 

\begin{figure}[hbtp]

\begin{lstlisting}[language=Python]
question = '''
You are implementing a cache using the MESI cache coherency protocol. The core above the cache sends a store request to the cache. What state(s) would the cache need to be in to allow the store to hit in the cache and no request be sent on the bus to other caches or memory?

Choose exactly one answer from the following:
A. S (Shared)
B. I (Invalid) or S (Shared)
C. E (Exclusive)
D. M (Modified) or E (Exclusive)
'''
reference_answer = 'D'
\end{lstlisting}
\caption{Sample Multiple-Choice question}\label{MCQuestion}
\end{figure}

The questions in both categories spanned various topics of relevance to the design team, like Digital Design, Instruction Set Architecture, Processor Microarchitecture, VHDL Syntax, etc. Questions generated by one expert were examined by another expert, edited if necessary, and rated for relevance to the project and for difficulty. Inappropriate questions were dropped as also were questions that were too long for our context window (8192 tokens). At the end of this process we had 80 Code Explanation questions and 263 Multiple-Choice questions.

\subsection{Extended Pretraining}

We perform an extended pretraining of a base Granite foundation model\cite{granite-code} to improve its performance at explaining and generating VHDL code. The base Granite model was trained in two separate phases using two data distributions~\cite{granite-code}: (1) 3 trillion tokens of code covering 116 programming languages; (2) 500 billion tokens of code/natural language for the model to improve in the areas of reasoning and problem solving. Our extended training further tunes the model with high-quality VHDL data comprising both code and documents. The dataset was curated and filtered by subject matter experts (SMEs) as mentioned in Section \ref{doc_acquisition}. In order to avoid catastrophic forgetting we also included replay tokens sampled from the phase-2 training data distribution. We detail the data distribution and token count in Table~\ref{data_distribution}.

\begin{table}[h!]
\centering
\caption{Token distribution used in extended pretraining}
\begin{tabular} {|l | r|}
\hline\hline
\multicolumn{1}{|c|}{Token Type} & \multicolumn{1}{c|}{Token Count} \\
\hline
Code in VHDL and variants & $162$M \\
Code in other languages & $168$M \\
Documents & $14.6$M \\
\hline
\end{tabular}
\label{data_distribution}
\end{table}

The model was trained using a fork of NVIDIA's Megatron framework. The data was packed to fill up the full $8192$ context window of the model. The documents were separated from each other by [EOS] tokens, and the $4$-D attention mask definition prevents cross-contamination. The global batch size was $512$, the learning rate was changed using a cosine annealing decay with $25$ warm-up steps and a maximum value of $5\times10^{-5}$. The model was trained on $420$ steps (i.e. $1.76$B tokens) on one server with 8 NVIDIA H100 GPUs installed in our secure computing cluster.

\subsection{Instruction tuning}
After the extended pretraining, the model was further instruct-tuned on $1.1$M documents that included various datasets selected to improve instruction following capabilities of the model. Details of this dataset may be found in~\cite{granite-code}. For instruct-tuning we used a batch size of $128$, a learning rate of $5\times10^{-5}$, with a warm-up phase of $200$ steps, and a cosine decay on $20,000$ steps (i.e. $2.5$ epochs). We used $32$ NVIDIA A100 GPUs with $80$GiB memory.

\subsection{Model Merging}
Ideally, every model we evaluate must be instruction-tuned so that it can follow instructions in the prompt well. However, the process of instruction tuning is quite expensive with each run requiring 18 hours despite the use of 32 GPUs. To drastically reduce the time to get an instruction-tuned model, we experimented with the model-merging technique\cite{ties-merging}\cite{chip-align}. We took an available instruction-tuned version of the base model\cite{granite-code}, and merged its weights with the weights of the EPT model using the spherical linear interpolation (SLERP) technique. This is not difficult to do because both models were derived from the same base LLM. 

\subsection{Model Evaluation}\label{eval_section}
Evaluation of the model needs to be done at different phases in the model development. The technique used to evaluate is different in each of the phases.
\begin{enumerate}
\item During the hyperparameter determination phase, the model evaluation was done simply by monitoring the loss function.
\item During the model training phase, model evaluation was done by using a validation set from the Multiple-Choice test set. We also guarded against catastrophic forgetting by monitoring the performance on standard benchmarks, e.g. HumanEval\cite{chen2021evaluatinglargelanguagemodels}. The results of these tests were used to guide the sampling parameters for new training runs. We discovered that the mix of three sets of data was most important: replay data, code, and documents. Too much replay data slows down learning of new material, while too much code in comparison with documents slows down knowledge acquisition. In the end, we settled on 40\% of the data seen during training being replay data from base model pretraining, 28\% code, and 32\% documents. The online evaluation was performed on a different, less expensive cluster consisting of a pair of L40 GPUs and overlapped with the training on the H100 cluster.
\item A small set of checkpoints was selected and tested offline using the Code Explanation test set. Offline evaluation has less strict resource constraints and hence can be longer. Explanations produced by the models were evaluated by SMEs. Each expert was assigned a set of questions for evaluation. The experts were required to evaluate each question on a 5-point Likert scale in 4 categories: correctness, completeness, compactness, and consistency. A rubric was provided to ensure consistency in evaluations. A screenshot of the tool we created to help in the evaluation is shown in Fig. \ref{eval_ss}.
\begin{figure}[hbtp]
    \centering
    \includegraphics[width=0.85\linewidth]{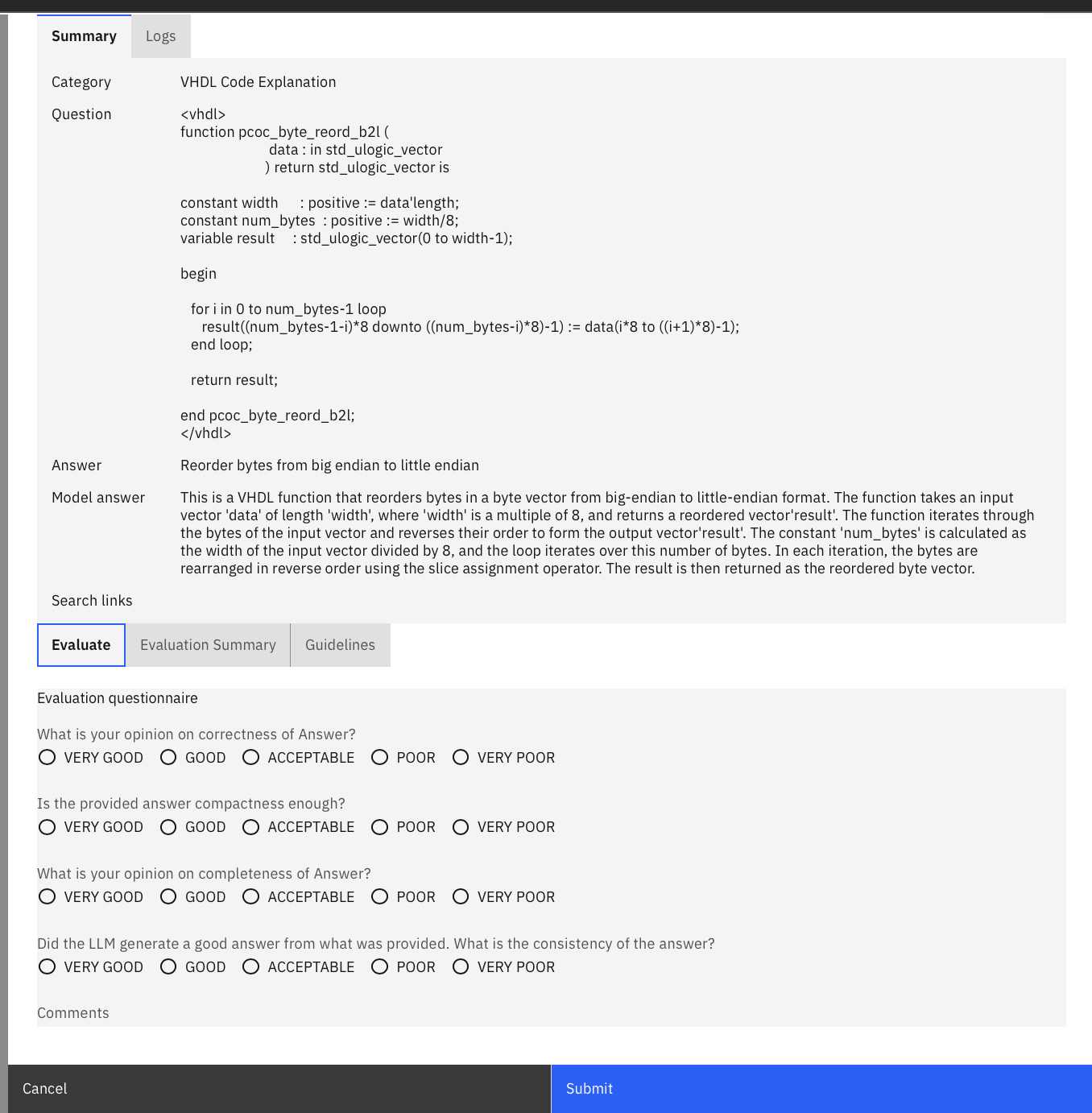}
    \caption{Screenshot of SME Evaluation Platform}
    \label{eval_ss}
\end{figure}

With new models being developed constantly, we found that it was getting difficult to get evaluations from SMEs quickly and reliably. This led us to developing an LLM-as-a-judge to perform the evaluations. The prompt that we used is shown in Fig. \ref{judge_prompt}. Three different LLMs were tried out for this purpose.
\begin{figure}[hbtp]
\begin{lstlisting}[language=Python]
instruction = '''
Below is a piece of VHDL code, a reference explanation produced by an expert that describes what the code is doing, and a model explanation that was produced by an AI model. You are required to evaluate whether the model explanation is similar to the reference explanation using a point system described below.
- Give 5 points if the model explanation captures all the concepts expressed in the reference explanation
- Give 4 points if the model explanation captures most of the concepts expressed in the reference explanation
- Give 3 points if the model explanation captures the essential concepts expressed in the reference explanation
- Give 2 points if the model explanation captures a minority of concepts expressed in the reference explanation
- Give 1 point if the model explanation captures almost none of the concepts expressed in the reference explanation
Here is the package consisting of the input code, the reference explanation, and the model explanation:
{eval_pkg}
After examining the package:
- Briefly justify your score, up to 100 words.
- You must provide the score exactly using the following format: "VHDL Code Explanation Score: <points>."
'''
\end{lstlisting}
\caption{Sample prompt template used for LLM-as-a-judge}\label{judge_prompt}
\end{figure}
\item With the final model deployed in the field, we have also enabled the gathering of telemetry data and user feedback to identify potential areas for further improvement of the model through continued pretraining, post-training and alignment techniques. This is a key requirement for adoption and integration of LLM-aided tools in production design flows.
\end{enumerate}

\subsection{Beam Search}\label{beam_search_section}
Beam Search\cite{beam-search} is a common technique to explore the search space in a resource-efficient way by keeping a limited number of candidate tokens at any point, rather than the single “best” token. Beam search is a compromise between the efficiency of greedy search and the optimality of exhaustive search. It essentially attempts to identify \textit{token paths} that have low cost rather than \textit{tokens} that have the lowest cost. While the Code Explanation task is a candidate for such an exploration, the Multiple-Choice task, as we have currently formulated it is not, because of the limited length of the output sequence. For the Code Explanation task, we use the best judge chosen from our LLM-as-a-judge experiments to evaluate the model's performance for different beam widths. In the future, we plan to exploit different inference (test-time) scaling techniques\cite{llm-monkeys}\cite{welleck2024from} as the customized LLM gets deployed broadly across the hardware design flows.

\section{Results}\label{Results}

We performed two rounds of extended pretraining. The second round improved on the first round by changing the sampling ratios between code tokens and document tokens. In Table \ref{CE_results}, which shows the \begin{table}[h!]
\centering
\caption{Results on Code Explanation Test Set}
\label{CE_results}
\begin{tabular} {|c|c|c|c|c|c|c|}
\hline\hline
\multicolumn{1}{|c}{} &
\multicolumn{4}{|c}{Scores} &
\multicolumn{1}{|c}{SME} &
\multicolumn{1}{|c|}{Norm.J2}
\\
\hline
\multicolumn{1}{|c|}{Version}& 
\multicolumn{1}{c|}{SME}&
\multicolumn{1}{c|}{Judge 1}&
\multicolumn{1}{c|}{Judge 2}&
\multicolumn{1}{c|}{Judge 3}&
\multicolumn{1}{c|}{\%score}&
\multicolumn{1}{c|}{\%score}
\\
\hline
Base & 2.73 & 3.76 & 3.19 & 2.86 & 43 & 43\\
EPT1.1 & 3.25 & 4.14 & 3.35 & 3.06 & 56 & 53\\
EPT1.2 & 3.25 & 4.10 & 3.33 & 3.09 & 56 & 51\\
EPT2.1 & 3.60 & 4.36 & 3.49 & 3.30 & 65 & 60\\
EPT2.2 & 3.75 & 4.59 & 3.51 & 3.25 & 69 & 62\\
EPT2.3 & 3.55 & 4.49 & 3.55 & 3.36 & 64 & 64\\
\hline
\end{tabular}
\end{table}
performance of various models on the Code Explanation test set, the first row refers to the base model. The next two rows refer to the performance on the first round of extended pretraining, while the last three rows refer to the second round of extended pretraining. The models were judged by subject matter experts. The base model had a rating of 2.73 on the Likert scale, which is below the “acceptable” level of 3. It is clear that both rounds of pretraining improved on the performance of the base model. The second pretraining did considerably better with a score of 3.55-3.75 compared to the first round of pretraining, which had a rating of 3.25.

The additional columns in Table \ref{CE_results} show the scores determined by the three LLMs-as-judges that were described in Section \ref{eval_section}. It shows that all three judges as well as the SMEs order the models in a similar way, even though their scales differ slightly. The degree of correlation between the judge scores and the SME scores was 0.99 for Judge 1, 0.95 for Judge 2 and 0.93 for Judge 3. While the judges rated the models in a remarkably similar way as the SMEs, the scores on the individual questions had a weaker correlation of around 0.55, with Judges 2 and 3 doing better than Judge 1. We feel this may be due to the fact that multiple SMEs were involved and their evaluations may not have been consistent. For further experimentation, we used a normalized judge score by taking the score provided by Judge 2 and scaling it to be comparable to the SME score. The percent score is simply computed as $ 1/4 * (normalized\_score -1)$. This score is shown in the last column of Table \ref{CE_results} as Norm J2 \% score.

The results of instruction-tuning on the best pretrained model, EPT2.2, are shown in Table \ref{IT}. Also shown in the table is the performance of a merged model, derived by using SLERP between EPT2.2 and instruction-tuned version of the base model\cite{granite-code}.  
As evident from the table, model merging results in improved performance, though a full instruction-tuning is able to do even better at a much higher cost in hardware resources.
\begin{table}[h!]
\centering
\caption{Normalized scores for instruction tuned models}
\label{IT}
\begin{tabular} {|r|c|c|c|}
\hline\hline
\multicolumn{1}{|c|}{Model} & 
\multicolumn{1}{c|}{Normalized score} &
\multicolumn{1}{c|}{\% score} &
\multicolumn{1}{c|}{H/w resources}
\\
\hline
Current Base & 2.73 & 43 &\\
EPT2.2 & 3.47 & 62 & 25 hrs on 8xH100\\
Model merged & 3.67 & 67 & 10 mins on 2xL40S \\
Instruction-tuned & 3.84 & 71 & 18 hrs on 32xA100\\
\hline
\end{tabular}
\end{table}

We have deployed the instruction-tuned model for use by logic designers. In our first implementation designed mainly to get feedback on the usefulness of the tool, the user interacts with the model through a simple interface where the user pastes a piece of code and the model returns an explanation. The user is requested to provide feedback, both in the form of thumbs-up/thumbs-down, as well as detailed written comments about the strong and weak points of the code explanation. It is interesting to note that users appear to be generally reluctant to provide feedback, with less than 50\% of those who try the tool, providing useful feedback. Currently the thumbs-up/thumbs-down ratio is running at about 70\% to 30\%. The written feedback has already allowed us to improve the performance of the model simply through prompt engineering. Further improvements are targeted.

New, more capable models are being developed in the AI world at a relentless pace. The incorporation of reasoning capability in new models has resulted in breakthrough performance improvements in many use cases. Table \ref{CE_new_results} indicates the \begin{table}[h!]
\centering
\caption{Normalized scores for new base models}
\label{CE_new_results}
\begin{tabular} {|r|c|c|c|}
\hline\hline
\multicolumn{1}{|c|}{Model} & 
\multicolumn{1}{|c|}{Parameters} & 
\multicolumn{1}{c|}{Normalized score} &
\multicolumn{1}{c|}{\% score} 
\\
\hline
Current Base & 20B &  2.73 & 43 \\
New Base 1 & 8B & 3.67 & 67 \\
New Base 2 & 8B & 4.12 & 78 \\
New Base 3 & 14B & 4.38 & 85 \\
\hline
\end{tabular}
\end{table}
potential for high performance improvement we could see even for VHDL Code Explanation. The base performance of these models on our Code Explanation test set improves from 43\% for our current base model to 78\% for a future model with only 8B parameters, and improves to 85\% for a model with 14B parameters. Extended pretraining or targeted fine-tuning of these new foundation models could yield a model that may even surpass SMEs in their code explanation capability.

The results from beam search on Code Explanation are shown in Table \ref{beam_search}. 
\begin{table}[h!]
\centering
\caption{Effect of Beam Search on Instruction-Tuned Model}
\label{beam_search}
\begin{tabular} {|c|c|c|c|}
\hline\hline
\multicolumn{1}{|c|}{Beam Width} & 
\multicolumn{1}{c|}{Normalized score} &
\multicolumn{1}{c|}{\% score} 
\\
\hline
1 & 3.84 & 71 \\
2 & 4.09 & 77 \\
3 & 4.01 & 75 \\
4 & 3.81 & 70 \\
5 & 4.09 & 77 \\
\hline
\end{tabular}
\end{table}
Because of memory constraints we limited our experiments to a beam width up to 5. While there is potential for improvement using beam search, it is hard to determine a priori the correct beam width for a problem, because the performance does not improve  monotonically with beam width.  However, this experiment did yield a considerably better model as evaluated by our scaled judge, improving from 71\% for a beam width of 1 to 77\% for beam widths of 2 or 5.

The results from administering the Multiple-Choice test to various models are shown in Table \ref{MC_results}.  
\begin{table}[h!]
\centering
\caption{Results on Multiple-Choice Test Set}
\label{MC_results}
\begin{tabular} {|r|c|c|c|}
\hline\hline
\multicolumn{1}{|c|}{Model} & 
\multicolumn{1}{|c|}{Parameters} & 
\multicolumn{1}{c|}{MC Test Accuracy} 
\\
\hline
Current Base & 20B &  0.34 \\
After EPT & 20B &  0.36 \\
After IT & 20B &  0.36 \\
New Base 1 & 8B & 0.39 \\
New Base 2 & 8B & 0.39 \\
New Base 3 & 14B & 0.61 \\
\hline
\end{tabular}
\end{table}
Here are some observations:
\begin{itemize} 
\item There is only a marginal increase, from 34\% to 36\%, in the performance of the model on our Multiple-Choice Test after extended pretraining and instruction-tuning (first three rows).
\item The data on the last three rows of the table indicate the capability of some of the new models on the horizon. They lay to rest any fears that the Multiple-Choice test set is impossibly hard. The base capability of the new 8B parameter models  surpass the capability of the pretrained and instruction-tuned 20B models. Newer base models achieve an even higher 61\% before adaptive pretraining and instruction-tuning, though with a larger number of parameters (14B). All three models differ from our original base model in their incorporation of reasoning capabilities. 
\end{itemize}
We believe that advances in the reasoning ability of future models will not only make the performance of these and their derivative models do well on such Multiple-Choice questions, but also on tasks of special importance to the microprocessor and hardware community, such as debugging and verification.

\section{Discussion}\label{Discussion}
This paper outlined our process of creating an AI assistant for understanding VHDL code in an industrial setting. Our requirements, specifically the need to exploit legacy assets, the need to perform training and inferencing in a secure environment, the need to be sensitive about the provenance and permissions associated with external data, and the need to be sensitive about availing the services of human experts, while different from common settings, are not unique. These requirements add cost and schedule constraints to large-scale industrial projects, but are not insurmountable. 

We are in the process of deploying the model to help our new designers get up to speed on existing VHDL designs, often used as a base for new designs or new units in a design. We are also planning further improvements to the model by
\begin{itemize}
\item Further fine-tuning of domains where the model is weak
\item Expanding the range of tasks which the model could be tackling
\item Repeating the extended pretraining and fine-tuning processes with newer reasoning models as base, and
\item Enhancing the model capability using newer developments in Generative AI, especially around reasoning and inference scaling.
\end{itemize}

We are confident that we will soon be developing models which will approach the performance of experts, perhaps even surpassing them in consistency of performance.

The area of high-performance processor design, with its use of leading-edge technologies and microarchitectural techniques, has traditionally had longer design cycles than other types of design. The deployment of LLMs to aid the process of specifying, verifying, and testing such designs will go a long way in reducing these design cycles and spur innovation in the processor world at an even faster pace than today.

\section*{Acknowledgment}
The authors thank the data and model factory team, particularly Maroun Touma, Hima Patel, Bhatta Bhattacharjee, and Nirmit Desai for help with the preprocessing of data, and Rameswar Panda for the base model replay data. Thanks also to Apoorve Mohan, Rick Welp and their teams for help with the infrastructure. This project would not have been possible without the help of dozens of logic design experts who are critical to the design of IBM's high-end P-series and Z-series processors. Special thanks also to Matt Logelin who used all his charm and wiles to coax and cajole these experts. Finally, we would like to thank Doc Vaidhyanathan, Susan Cohen, Xuan Liu, and Hillery Hunter, whose unwavering support and guidance made this work possible.

\bibliographystyle{IEEEtran}

\end{document}